# NON-TOXIC FABRICATION OF FLUORESCENT CARBON NANOPARTICLES FROM MEDICINAL PLANTS/SOURCES WITH THEIR ANTIOXIDANT ASSAY

Parul Singh[1], Aniruddha Dan[2], Padma Priya Kannan[3], Mukesh Dhanka[2], Dhiraj Bhatia[2], Jhuma Saha[1*]

[1] Department of Electrical Engineering, Indian Institute of Gandhinagar

[2] Department of Biological Science and Engineering, Indian Institute of Gandhinagar

[3] Department of Materials Engineering, Indian Institute of Gandhinagar

[1*]correspondence to email: jhuma.saha@iitgn.ac.in

ABSTRACT: This research work showcases a non-toxic approach to synthesize carbon nanoparticles (CNPs) from various medicinal plants namely Syzygium cumini, Holy basil, Azadirachta indica A, Psidium guajava, Mangifera indica, and Bergera koenigii using microwave approach. The optical, morphological, structural, and functional properties of obtained CNPs from all mentioned sources were investigated using UV-Vis, Scanning electron microscopy (SEM), Fourier transform infrared spectrophotometry (FTIR), dynamic light scattering (DLS), zeta potential tests and X-ray diffraction (XRD). With great water dispersibility, and photostability all the medicinal sources chosen yielded in bright red fluorescent nanoparticles under exposure to UV light, thereby giving a significant peak around 650 nm recorded in absorption spectrum. Antoxidant assay was performed on all these six different plant-derived nanoparticles with two different concentrations and all have exhibited excellent free radical (DPPH) scavenging activity, proving their role as antioxidants. This further opens up doors for various other plant and biomedical applications to be targeted using these CNPs.

KEYWORDS: carbon; medicinal; non-toxic; spectroscopy; scavenging

INTRODUCTION:
In recent years, carbon-based nanomaterials have received widespread acclaim from around the world, with various interdisciplinary research groups stepping into the realm of nanoscale to tailor the properties on an atomic level. Although semiconductor devices dominate the market presently, the exponentially growing awareness to design environmentally friendly materials has driven the scientific community to look for alternatives and the development of carbon dots show great promise to help fill the void.
Carbon is one of the most abundant materials on Earth and forms the basic building block of all living matter. Carbon exists in its allotropic form, crystalline diamond and graphite in its amorphous coal, charcoal, and lampblack. It is challenging to tailor and design a material with appropriate bandgap tunability and effective photoluminescence. It thus becomes integral to study materials at the nanoscale. Carbon Dots are classified into three types, namely Graphene quantum dots, Carbon nanodots, and Polymer Dots. Carbon dots have unique electrical and optoelectronic properties, an inherent bandgap, and high dispersibility in a variety of solvents. This class of materials display excellent behavior in biological environments, non-toxic to different cell lines even at high concentrations thus making them highly biocompatible [1,2]. The presence of abundant functional groups on the these make it more favorable to tailor its properties through surface modifications with different molecules or polymers, therefore customizing its properties. A polymer coating can control optical properties such as emission wavelength, lifetime, color while metal coatings can provide the material with conductivity and enhanced biological activity.

Along with easy surface functionalization, the new entrant carbon dots with water solubility, photostability, and non-toxic composition have shown a strong potential in biomedical engineering (drug delivery, antibacterial effect, bio-imaging), metal-ion sensing, optoelectronic devices, in catalysis and a whole new plethora of applications [3-6].

With the world population increasing rapidly, the impact on the agricultural sector has been immense. The use of pesticides and fertilizers has imposed a severe threat to the global ecosystem. The attractive properties of nanomaterials are being investigated to alleviate these challenges. Additionally, the low cost of production, simple synthesis procedure and versatility in applications make this field of research rewarding. The carbon dots have shown the ability to penetrate through the plant cells helping in the transport of integral nutrients and are thus being considered as effective alternatives to conventional fertilizers. Due to their biocompatibility carbon dots have also found their applications in the agricultural sector. CDs have shown to improve plant health and agricultural yield. The light-harvesting potential enables the CDs to act as artificial photo-antennae, absorbing UV light and transforming them into blue-violet or red light thereby increasing the rate of photosynthesis of the leaves and promoting plant growth [7,8]. Although there are multiple routes through which carbon dots can be synthesized the most infamous one remains the green synthesis. Green carbon nanoparticles (CNPs) are synthesized through natural resources and biomass including fruits, vegetables, plant leaves, and human derivatives. The exclusive benefit being the chemical-free synthesis, favorable operating conditions, with small particle size, tunable photoluminescence (PL) and can be easily conjugated with biomolecules [9,10]. CNPs synthesized from chemical compounds are impaired with low quantum yield, hazardous precursors and include huge capital costs. The biomass-derived dots on the other hand can be synthesized on large scale while being sustainable, and are self-stabilized owing to the proteins, flavonoids, and the phytochemicals present in the parent source [11-13].

The synthesis of CNPs can be achieved through two strategies, namely top down and bottom up. In the top-down synthesis large carbon precursor molecules are broken down into nanoscale particles, although control over the size and properties can be accomplished through bottom-up synthesis, where smaller units such as organic molecules are assembled and transformed to CNPs. The techniques in this domain include:- microwave irradiation, hydrothermal and solvothermal synthesis. Both the hydrothermal and solvothermal synthesis are carried out in high temperature and pressure conditions in an autoclave. They are the most common route adopted by research groups with low experimental requirements and achieving CNPs with good quantum yield [14,15]. Recent studies show that a variety of applications have been targeted using the green synthesis approach to produce CNPs from natural resources, thus making them a suitable candidate and an emerging alternative to chemical-induced nanoparticles [16-19]. Significant research is underway to comprehend the actual mechanism behind the formation of CNPs through green synthesis. Additionally, there is an exploration into the synthesis of charged carbon dots. Positively charged CNPs have demonstrated bactericidal properties by causing damage to the cell surface. However, acquiring stable charged CNPs presents a substantial challenge, necessitating further research and innovative solutions to address issues related to their long-term stability and performance. In this work, the microwave approach is highlighted and promoted because the process is easy to handle, cost-effective, eco-friendly, and does not require long hours of heating the substrates. Microwave irradiation helps synthesize carbon dots in a matter of minutes, where the precursor is irradiated with electromagnetic radiations to form CNPs.

MATERIALS AND METHODS:

Six medicinal plants are used in this study, namely Mangifera indica (commonly known as Mango, so nanoparticles obtained from this source is addressed here as MNPs {*Figure 1A*}), Azadirachta indica (commonly called Neem, thus addressed as NNPs {*Figure 1B*}), Psidium guajava (commonly called Guava, thus GNPs {*Figure 1C*}), Holy basil (commonly called Tulsi, so TNPs {*Figure 1D*}), Bergera koenigii (commonly called as Curry Leaves, thus CLNPs {*Figure 1E*}), and Syzygium cumini (commonly called Jamun, thus JNPs here {*Figure 1F*}). Fresh leaves of these six sources mentioned in the figure 1 were plucked from the IIT Gandhinagar campus. The freshly plucked leaves were thoroughly washed and then shade dried. The powder obtained after grinding the shade dried leaves was then mixed with ethanol. After few hours of constant magnetic stirring, the solution was centrifuged at 6000 rpm for ten minutes, and then the supernatant was extracted. This was followed by five minutes of intermittent stirring in a domestic microwave (800W). The resulting slurry mixture was then mixed with a small amount of DI water if necesaary. The residue was then syringe filtered using 0.22μm syringe filter, and was allowed to dry at 60°C for a few hours in hot oven to obtain dried powder.

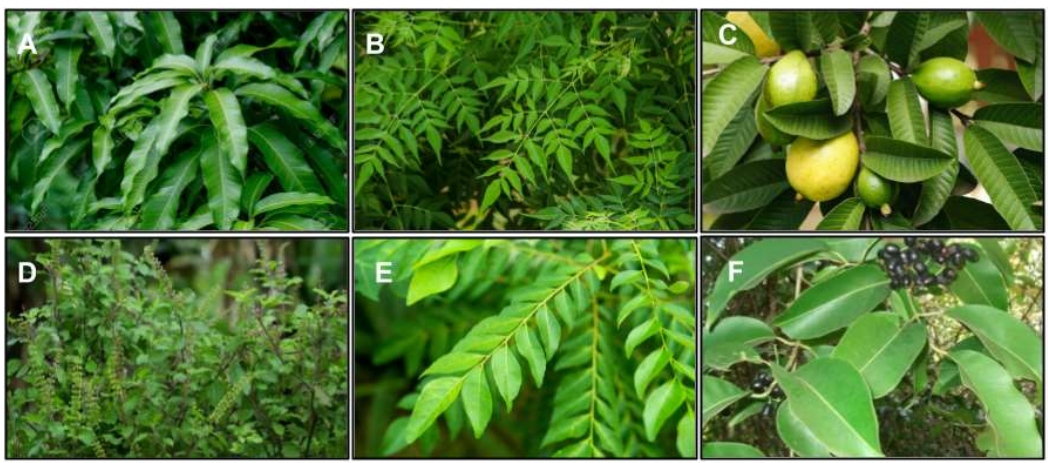

*Figure 1: Fresh leaves of Mangifera indica [A], Azadirachta indica A [B], Psidium guajava [C], Holy basil [D], Bergera koenigii [E], and Syzygium cumini [F] were plucked from the IITGN campus and processed for synthesizing CNPs.*

*FOR ANTIOXIDANT ASSAY*

DPPH assay was conducted to investigate the antioxidant activity of the synthesized plant derived CNPs. DPPH scavenging percentage was determined based on the absorbance at 517 nm in the presence of various concentrations of CNPs [20]. Two different concentrations of CNPs (1mg/ml, 2mg/ml) were added to the DPPH-methanol solution (100 μM), incubated in dark condition at 37°C for a certain time period. Equal volume of DPPH solution and Methanol (100%) were regarded as control and blank. DPPH scavenging (%) was calculated by following equation mentioned below.

$$\text{Free radical scavenging (\%)} = \frac{\text{Absorbance of control} - \text{Absorbance of CNPs treated DPPH solution}}{\text{Absorbance of control}} \times 100$$

CHARACTERIZATION:

Using PerkinElmer Lambda 365, UV-Vis absorption spectra of the samples in an aqueous solution were obtained. The X-ray diffraction (XRD) pattern was obtained using the Rigaku SmartLab X-Ray diffractometer with Cu Kα (Kα1= 0.1541 nm) irradiation. Fourier transform infrared (FTIR) spectroscopy was used to investigate the functional group that was present in the synthesized NPs. Microstructure images of the material were obtained using field emission scanning electron microscopy (FE-SEM).

RESULTS AND DISCUSSION:

ANALYTICAL CHARACTERIZATION:

UV-Vis SPECTROSCOPY:

In order to investigate the optical properties exhibited by the nanoparticles, absorption spectra was recorded for each sample (Figure 2).

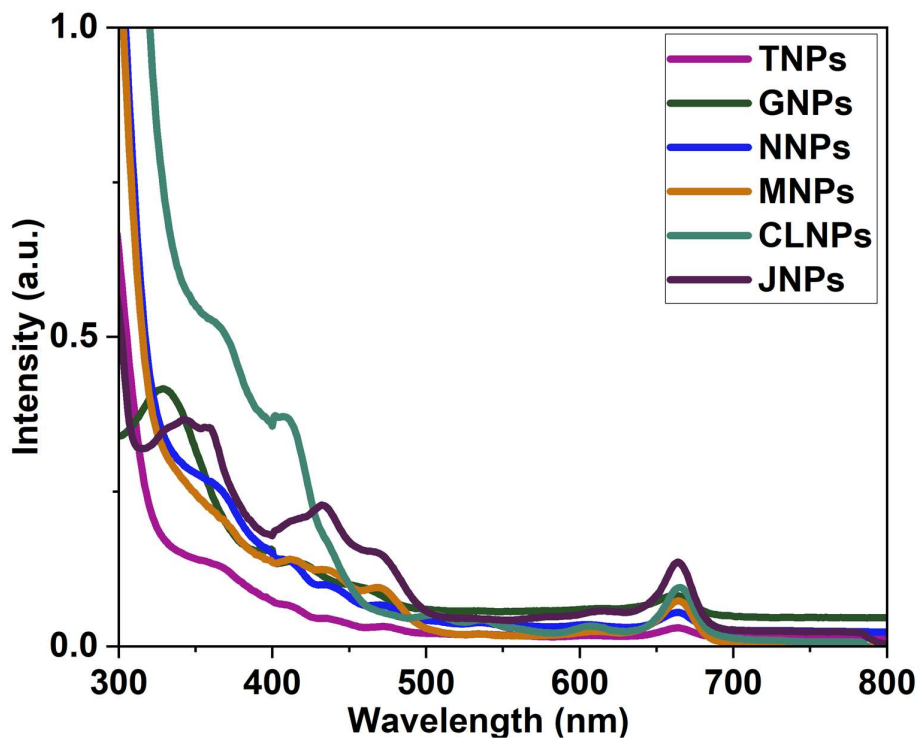

Figure 2: The absorption spectra of all the six sources is as shown. The figure also depicts the color of nanoparticle solution as green under the exposure of white light and bright red flouroscent under the exposure of UV light (inset).

For the case of TNPs, NNPs, CLNPs, and MNPs, peaks observed at 368 nm, 372 nm, 379 nm, and 365 nm respectively corresponds to π-π* transitions of the graphite $sp^2$ domains. Whereas the peaks recorded at 414 nm, 406 nm, 440 nm and 410 nm for CLNPs, NNPs, JNPs and MNPs respectively corresponds to n- π* transitions (C=N bonds) in the functional groups present. A peak at 337 nm and 352 nm for the case of GNPs and JNPs respectively might be attributed towards C=O strecthing. Also, one significant peak of absorption is observed for all the samples in the Near-infrared region (NIR) around 665 nm *(inset of figure 1)* corresponding to the red bright flouroscence exhibited by all the samples under the exposure of the UV light [21]. This red flouroscence showcases the huge potential of these derived CNPs in therapeutics, because of the tendency of this higher wavelength to site deep seated tissues at the area of interest.

SCANNING ELECTRON MICROSCOPY (SEM):

To investigate morphological properties such as size and shape of the nanoparticles, SEM was performed, and the particle size distribution was calculated for each case as shown in figure 3. All the samples possessed quasi-spherical shape as depicted in the figure 3. For MNPs and NNPs, the average size distribution of the nanoparticles was around 32 nm and 33 nm respectively *(figure 3. A, B)*. Similarly for GNPs and TNPs, the average size recorded is around 37 nm *(figure 3. C, D)*. Whereas for CLNPs *(figure 3 E)*, the average size is higher amongst others, that is around 42.5 nm and for JNPs *(figure 3 F)*, the size recorded is smallest amongst the lot, that is 23.5 nm. The diameters of the CNPs ranged from 10-60 nm which is a diverse range and average size for all samples recorded was less than 45 nm, and this can further be reduced further by optimising and controlling synthesis procedure. Further, the energy dispersive X-ray analysis concluded the presence of carbon, nitrogen and oxygen for all the samples.

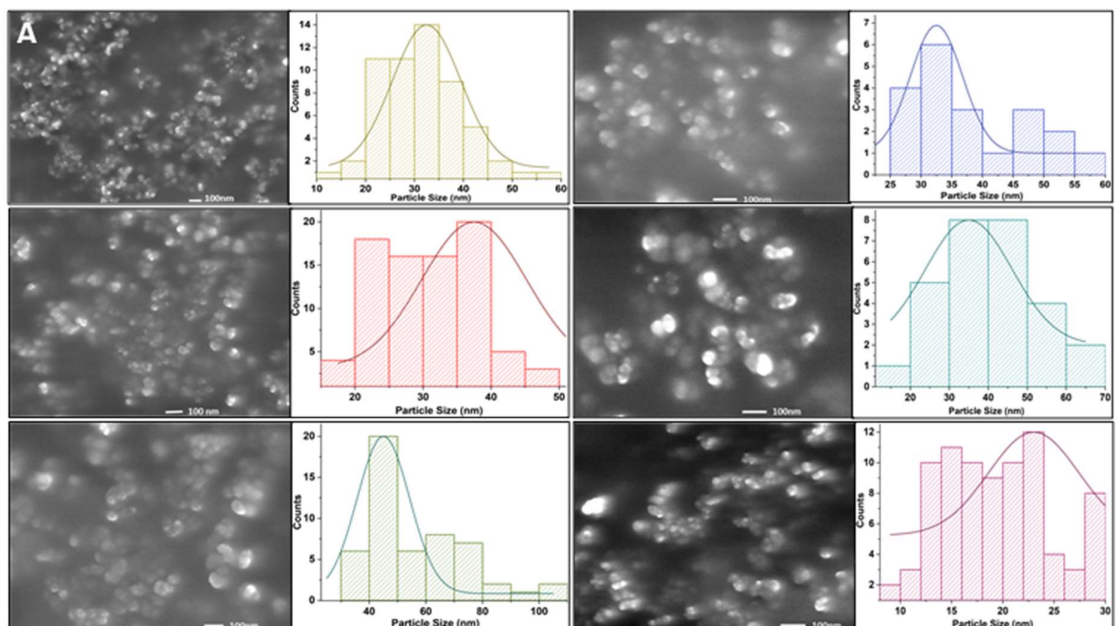

*Figure 3: The SEM image along with particle size distribution of (A) MNPs, (B) NNPs, (C) GNPs, (D) TNPs, (E) CLNPs, and (F) JNPs is as shown. The microscopy confirms the quasi-spherical nature of CNPs.*

FOURIER TRANSORM INFRA-RED SPECTROSCOPY (FTIR):

To investigate various chemical functional groups present the derives CNPs, FTIR was performed and is shown in Figure 4. The peaks observed at 1045 cm$^{-1}$ for GNPs *(Figure 4C)*, 1046 cm-1 for (MNPs, JNPs and NNPs) *(Figure 4A,B,F)*, and along with the peaks 1047 cm$^{-1}$ for the case of CLNPs *(Figure 4E)* attributes to the presence of C-O stretching vibrations.

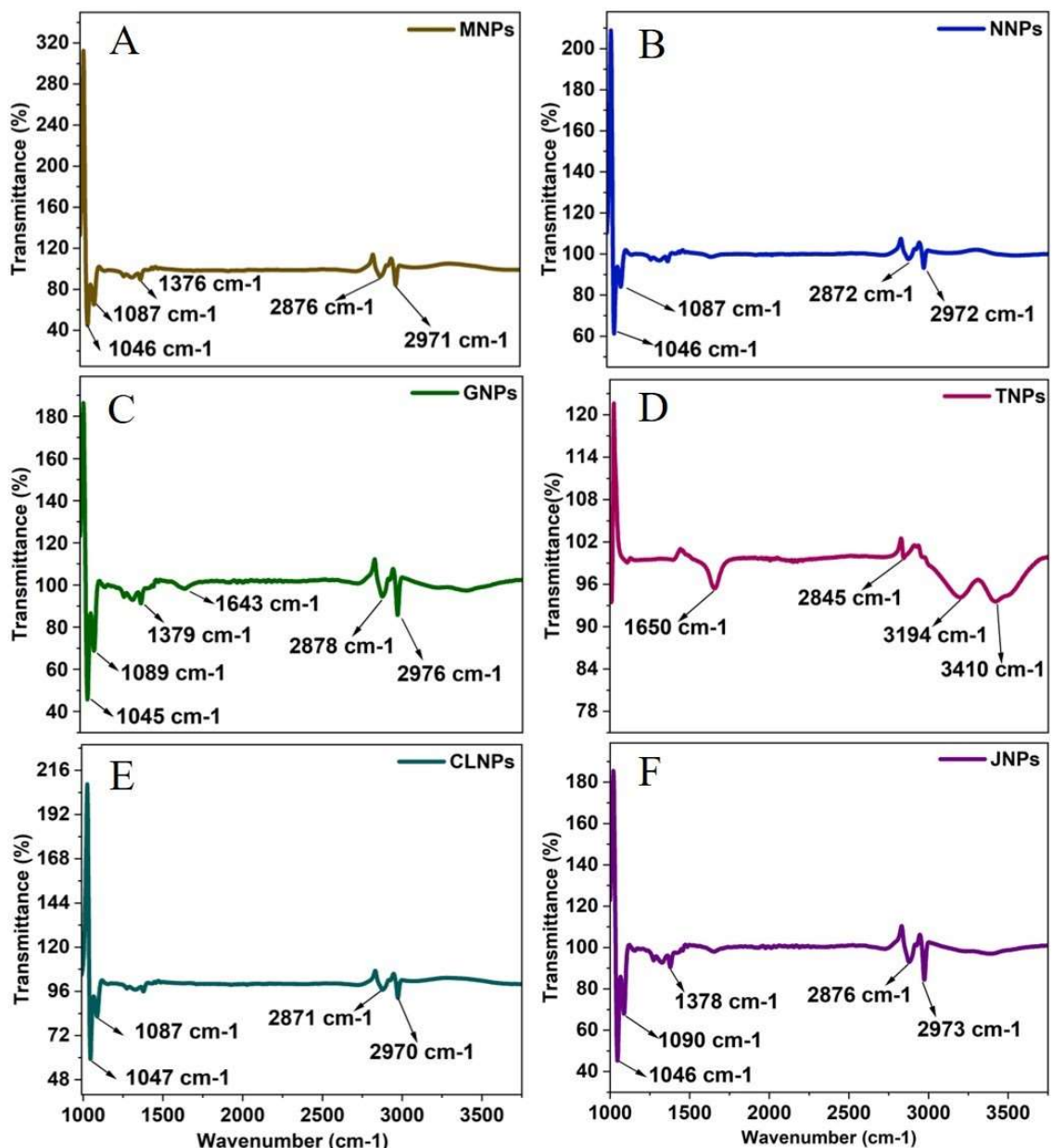

Figure 4: FTIR spectrum of (A) MNPs, (B) NNPs, (C) GNPs, (D) TNPs, (E) CLNPs, and (F) JNPs.

Whereas, the peaks observed at 1087 cm$^{-1}$ for MNPs and CNPs *(Figure 4A,E)*, and 1097 cm$^{-1}$ for JNPs and NNPs *(Figure 4B,F)*, along with 1089 cm$^{-1}$ for GNPs *(Figure 4C)* corresponds to C-N stretch (primary amines). While the peaks 1379 cm$^{-1}$ in GNPs, 1378 cm$^{-1}$ for JNPs, 2972 cm$^{-1}$ for NNPs, 2973 cm$^{-1}$ for JNPs, 1650 cm$^{-1}$ for TNPs, and 1376 cm$^{-1}$ for MNPs represnts bending vibrations of -OH groups. Further, the peaks observed in the range of (2870-2890) cm$^{-1}$ for samples MNPs (2876 cm$^{-1}$), JNPs (2876 cm$^{-1}$), NNPs (2872 cm$^{-1}$), and GNPs (2878 cm$^{-1}$) embarks the presence of methyl assymetric strectching of C-H. For the case of TNPs, the peaks observed at 3194 cm$^{-1}$, and 3410 cm$^{-1}$ shows conjugated ketones [22]. We observe the presence of chemical functional groups is almost similar in all the cases due to the presence of natural hydrocarbons in theese sources and also due to absence of any precursors used in the process.

X-RAY DIFFRACTION (XRD):

The crystalline nature of green synthesized NPs was investigated using XRD analysis using 2θ Bragg's angle range from 20° to 80° is as shown in Figure 5.

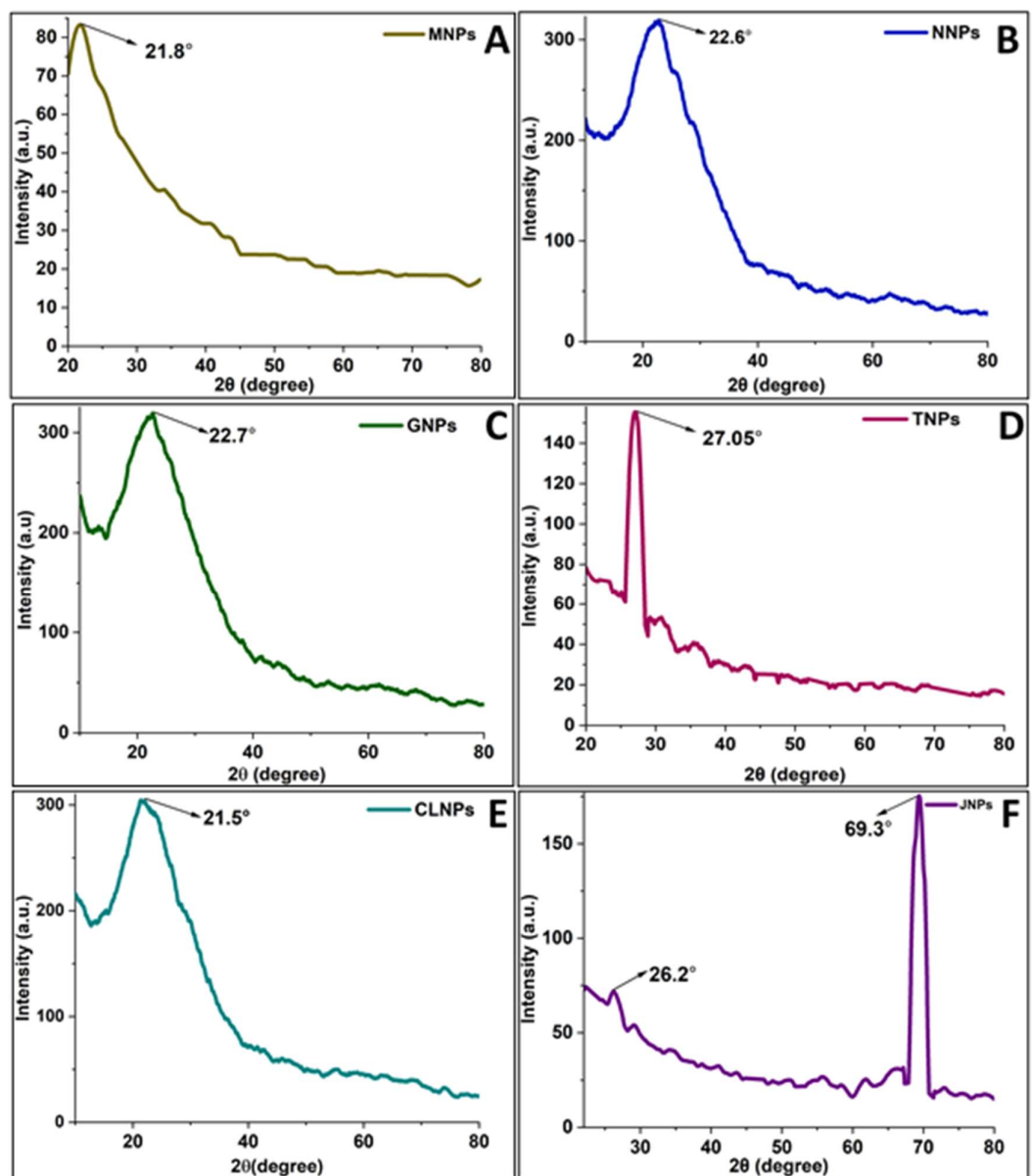

Figure 5: The XRD pattern of (A) MNPs, (B) NNPs, (C) GNPs, (D) TNPs, (E) CLNPs, and (F) JNPs.

The major broad peaks observed at 21.8° for MNPs *(Figure 5A)*, 22.6° for NNPs *(Figure 5B)*, 22.7° for GNPs *(Figure 5C)*, and 21.5° for CLNPs 21.8° for MNPs *(Figure 5E)* corresponding to (002) lattice plane with inter-layer spacing of 0.39 nm, 0.391 nm, 0.40 nm, and 0.41 nm respectively. This value is larger than the lattice plane of graphitic structure of (002) plane that is 0.34 nm, thereby showing poor degree of crystallinity. For samples JNPs, and TNPs *(Figure 5 D,F)* the sharp peak is observed at 26.2°, and 27.05° with spacing of 0.331 nm and 0.33 nm respectively showcasing better crystallinity structure at (111) plane [23]. Another significant peak is observed at 69.3° for JNPs with spacing of 0.14 nm for (400) plane with high crystallinity.

ANTIOXIDANT ASSAY

For examining the free radical scavenging activity of the synthesized plant derived CNPs, DPPH was selected as the test radical. Two different concentrations of each sample was prepared (5mg/ml and 10mg/ml) using DI water as solvent. These concentrations were then added to DPPH solution and scavenging was recorded after 5 minutes and after 30 minutes of the solution. Also, the figure 6 shows the transition of the DPPH solution from purple to yellowish, representing the scavenging of DPPH free radicals by the plant leaf extract derived CNPs, and this behaviour was common to all the samples. Then optical density values were recorded. Being an N-centered free radical, DPPH exhibits a characteristic peak at 517 nm.

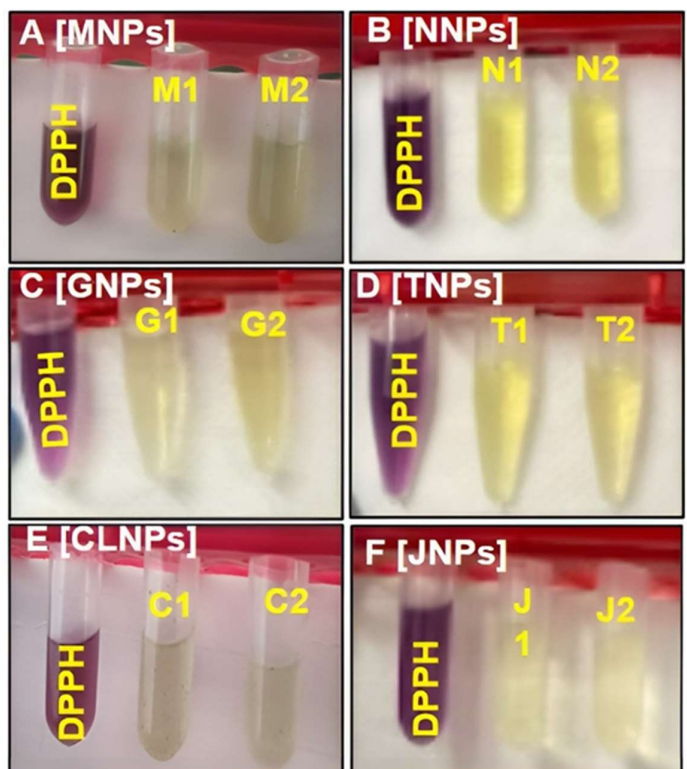

*Figure 6: Antioxidant assay of (A) MNPs, (B) NNPs, (C) GNPs, (D) TNPs, (E) CLNPs, and (F) JNPs with two concentrations (1 = 1mg/ml, and 2 = 2mg/ml).*

When DPPH comes in contact with free radical scavengers that is the derived CNPs, the absorbance decreases due to the formation of the DPPH complex [19]. The elimination of DPPH proves the anti-RNS ability. As shown in figure 7, the remaining DPPH (%) has gradually declined, as the concentration of all the CNPs has been increased for all samples.

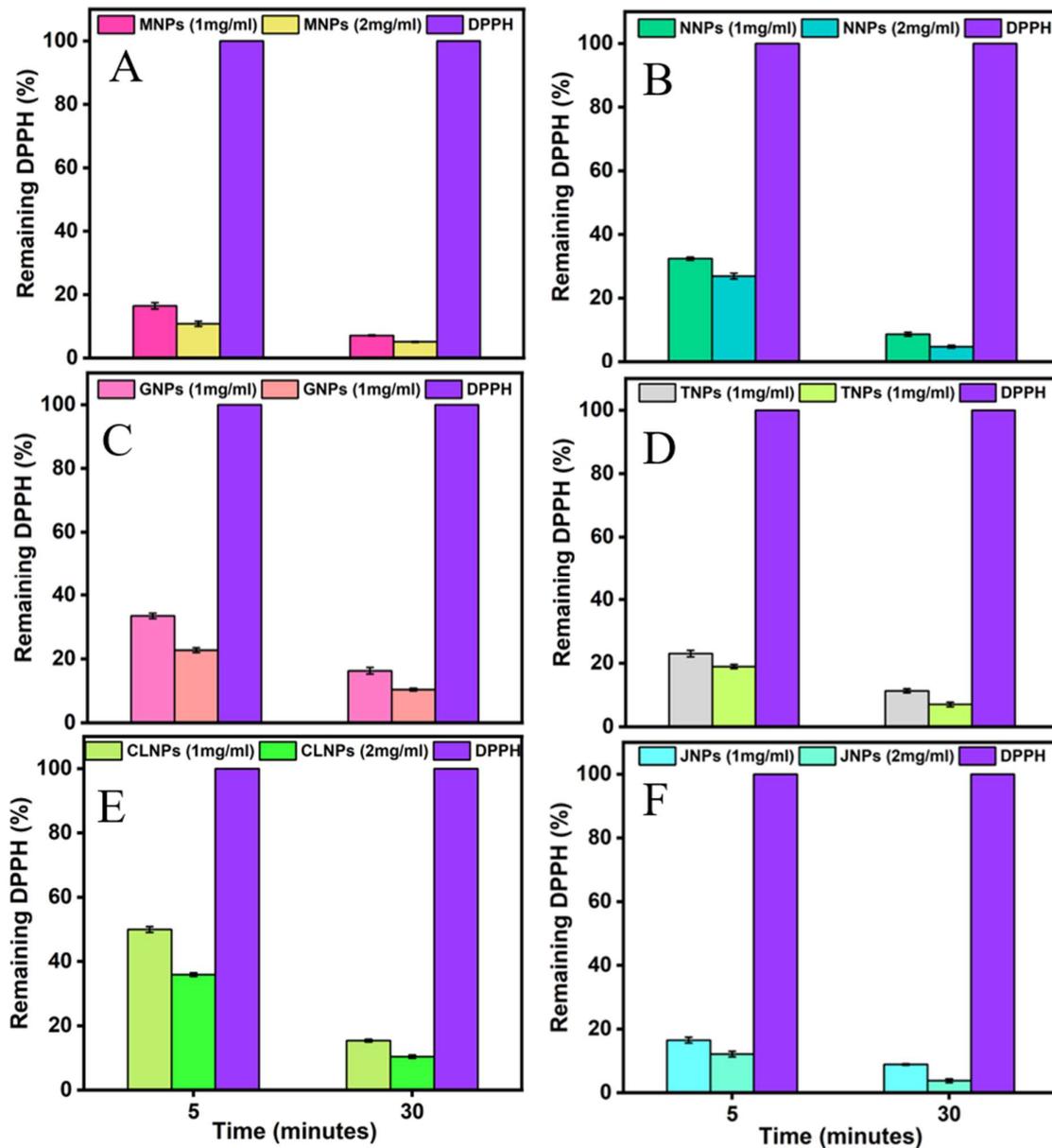

*Figure 7: Scavenging activity recorded after 5 and 30 minutes for (A) MNPs, (B) NNPs, (C) GNPs, (D) TNPs, (E) CLNPs, and (F) JNPs.*

Further, figure 7 depicts that, all of the six different plant leaf extract-derived CNPs have significant free radical scavenging activity, owing to the presence of several bioactive compounds in the leaves of these plants [24-35] and scavenging activity has been increased along with the increase in concentration of the respective CNPs. Among all fabricated CNPs, JNPs & MNPs scavenged the free radical (DPPH) in a very rapid manner compared to other plant leaf extract derived CNPs. The free radical scavenging activity of these plant leaf extract-derived CNPs are in order of- JNPs>MNPs>GNPs>NNPs>TNPs>CLNPs.

CONCLUSION:
This study demonstrates the efficient production of CNPs using plant extracts while examining their scavenging capabilities. The approach advocated here prioritizes the reduction of harmful chemicals, offering a superior alternative to current industry practices for nanoparticle synthesis. The selection of all six medicinal plants emphasizes their widespread availability, suggesting the potential for large-scale CNP production. Nanoparticles produced through green synthesis exhibit remarkable antioxidant properties, paving the way for exploring their potential in various tests, including anti-bacterial, anti-fungal, anti-cancer, and beyond. The vibrant red fluorescence when exposed to UV light highlights the potential of these nanoparticles as promising candidates for therapeutic applications such as bioimaging, drug delivery and much more. Further, the scope of these CNPs for plant health applications will also be tested and explored.

ACKNOWLEDGEMENT
We wholeheartedly thank the Central instrumentation facility (CIF) at the Indian Institute of Technology, Gandhinagar